\documentclass[twocolumn,article,amsmath,amssymb,aps]{revtex4}
\usepackage{graphicx}
\usepackage{epsfig}
\usepackage{multirow}
\usepackage{color}
\usepackage{cancel}
\begin{document}

\title{Some exact results for the particle number projected BCS approach 
of the isovector proton-neutron pairing}
\author{A. A. Raduta$^{1,2}$, M. I. Krivoruchenko$^{3,4}$, Amand Faessler$^{5}$}
\affiliation{$^{1}$Institute of Physics and Nuclear Engineering$\mathrm{,}$ P.O. Box MG06, Bucharest 077125, Romania\\
$^{2}$Academy of Romanian Scientists, 54 Splaiul Independentei, Bucharest 050094, Romania}

\affiliation{$^{3}$Institute for Theoretical and Experimental Physics$\mathrm{,}$ B. Cheremushkinskaya 25\\
117218 Moscow,\footnote{https://sites.google.com/site/saveitep/} Russia\\
$^{4}$Department of Nano-$\mathrm{,}$ Bio-$\mathrm{,}$ Information and Cognitive Technologies\\
Moscow Institute of Physics and Technology$\mathrm{,}$ 9 Institutskii per.\\
141700 Dolgoprudny$\mathrm{,}$ Moscow Region$\mathrm{,}$ Russia}
\affiliation{$^{5}$Institut f\"{u}r Theoretische Physik der Universit\"{a}t  
T\"{u}bingen, D-72076 T\"{u}bingen, Germany} 

\begin{abstract}

The mean values of a many-body  Hamiltonian  including a proton-neutron pairing term and  matrix elements of one-, two- and four-body operators within a basis of
particle number projected BCS states, are  analytically expressed in terms of a single function $Q(N)$ 
depending on the number of particles, $N$. The function $Q(N)$ is calculated using a recursion in $N$ 
in which the  shells and the BCS angles are kept the same for any step of iteration.
An illustrative example  is numerically considered in a restricted 
single particle space. Some specific features for the standard BCS, the projection after variation approach 
as well as for the variation after projection formalism, are pointed out. 
\end{abstract}
\maketitle 

\renewcommand{\theequation}{1.\arabic{equation}}
\setcounter{equation}{0}
\label{sec:level1}
\section{Introduction}
Short after the theory of superconductivity \cite{BCS,BOGO} showed up  it has been realized \cite{BMP} that such a formalism might work also for nuclear systems although the constituents number is relatively small. Many applications have been performed with a single constant for the interaction strength of various paired states. All calculations were based on the supposition that the particle number is conserved only in the average. Amazingly the pairing force and the emerging
seniority scheme has been introduced much earlier, by Racah \cite{Rac}.
A serious question arose, namely to what extent the particle number fluctuations affect some physical observables.
Attempting to answer such a question many authors used a projected  particle number formalism. 
Two distinct calculations have been employed. Occupation probabilities are determined first variationally with a standard BCS wave function 
and then the components of given number of nucleons are projected out. The resulting procedure is conventionally called the particle 
number projected BCS and as we said already the projection is performed after variation. 
The second set of calculations perform the variation after projection. 
The particle number projection has been first considered by Bayman in Ref.~\cite{Bay}. 
There the averages of the Hamiltonian, including the mean field and pairing terms, 
and of particle number operator are expressed as functions of some residuum integrals which where estimated by the saddle-point method. 
The saddle point approximately satisfies an equation which is similar to the particle number equation of the BCS formalism. 
Under these conditions the Euler-Lagrange equations obtained with a particle number projected variational 
state are identical to the standard BCS equations.
The projection procedure was improved in Refs.~\cite{DMP,MRR}. The residuum integrals were calculated by the saddle-point method 
with the integral path chosen such that it crosses the saddle-point on a line of steepest descent. 
Moreover two-dimensional recursion formulas for  the residuum integrals were provided. 
An extensive analysis of ordinary BCS, PBCS (projected BCS with projection after variation) and FBCS (variation after projection) 
is performed within a two levels pairing model~\cite{MRR}. 

Another feature  which was considered, referred to the centrifugal Coriolis anti-pairing effect.
The Coriolis interaction tends to decrease the pairing strength and at a critical angular frequency the gap equation has only the trivial solution. In Refs. \cite{MotVal,ChanVal} it is found that crossing the critical point the rotational energy exhibits a discontinuity which as a matter of fact, is not confirmed by the FBCS calculations of Ref.\cite{MRR}.   
Clearly the occupation probabilities emerging from a FBCS formalism are different from those associated with the ordinary BCS, the difference being function of the pairing interaction strength. The idea of particle number projection was extended to angular momentum. Indeed, the cranking model with a particle number projected wave function was considered \cite{MaRa} to investigate the backbending phenomena. The simultaneous projection of the particle number and angular momentum from a  pairing correlated many body system has been considered in Ref.
\cite{GoFa} for light nuclei.  
Many papers have been devoted to the issues mentioned above \cite{Ring} focusing on explaining some properties like, gap parameter, moment of inertia, spectroscopic factors, pairing versus nuclear deformation, and angular momentum. 

Although not as extensively, the Cooper pairs of one proton and one neutron have also been investigated \cite{Gos,Camiz,Wolt,Taka,Good1,Good2}. The results reported there demonstrate that the generalized Bogoliubov-Valatin (BV) transformation including $pp$, $nn$ and $pn$ pairing is  appropriate for treating the pairing correlations in a self-consistent way, in spite of  some earlier  pessimistic views on this issue \cite{Lan,Flow}. Note that in a generalized (BV) formalism both the total number of nucleons, the isospin third component (T3) and isospin (T) are 
not conserved. Therefore a simultaneous projection for all the three quantum numbers is necessary.
This type of projection has been considered by several authors both numerically and analytically
\cite{CheMutFa,RaMo}. The integrability of a $pn$ pairing model was treated in Refs.
\cite{Links,Duke} by different methods. Thus the pairing Hamiltonian introduced by Richardson in Ref.\cite{Richard} was considered in the context of the quantum inverse scattering method. It is proved that the model is integrable by constructing explicitly the conserved commuting operators.  The eigenvalues of these operators were determined in terms of  the Bethe ansatz and finally an expression for energy eigenvalues was possible \cite{Links}. A different method is applied, in Ref.\cite{Duke}, to the same Richardson model which includes isospin-symmetry breaking terms.

It seems that the common mathematical content with the above pairing correlated system is a serious ground for considering 
the pairing effect for other systems. 
Pairing forces acting among the quarks in two-color QCD matter lead
to the color superconductivity 
\cite{KOND91,KOND92}, as also confirmed by
simulations in lattice gauge theories \cite{HAND06}. 
Superconductivity of metallic nanoparticles is discussed in Refs. \cite{Black1,Black2,Del}.
Due to its paramount importance it is worth simplifying the formalism applied to other degrees of freedom following the successful path used for pairing of alike nucleons.

In this context the present paper considers the isovector $pn$ pairing interaction with the projected total number of particles. We aim at obtaining tractable equations for the residuum integrals and finally for the norms  and matrix elements of projected states.

This study is organized as follows. In Section II we study the factorization procedure applied to the exponent of a linear combination of the $su(2)$ algebra generators. The motivation for  this investigation is the fact that the BCS function could be obtained by transforming the particle vacuum state with such operator.  This is shown in Section III. The particle number projected function is described in Section IV. Analytical results of the average values of various interaction terms are given in Section V. Numerical results for a pairing Hamiltonian considered in a restricted single particle space are given in Section VI The final conclusions are collected in Section VII. 

\renewcommand{\theequation}{2.\arabic{equation}}
\setcounter{equation}{0}
\label{sec:level2}

\section{Factorization of the rotation operator} 


In the theory of superconductivity bi-linear forms of creation and
annihilation operators of fermions satisfy the commutation relations for the
generators of rotations. For this reason, we shall first derive some very useful, in the theory of superconductivity, algebraic
relations for the rotation operators. Although some of them are well known we present them for the sake of completeness.

In quantum mechanics, the rotation of wave function is given by real angle $%
\theta $ and the real unit vector $\mathbf{n}$ 
\begin{equation}
\Psi \rightarrow \Psi ^{\prime }=\exp (-i\theta \mathbf{nJ})\Psi , 
\label{rotoper}
\end{equation}
where $\mathbf{J}$ are the generators of rotation, given in Cartesian
coordinates. We are interested in various equivalent  representations of the rotation operators.
Concretely, we wish to establish a
connection between the rotation parameters $\theta ,\mathbf{n}$ from Eq. (\ref{rotoper}) and those denoted by $\alpha
_{\mu }$ and $\beta _{\mu }$ ($\mu =\pm ,0$)  which define two independent factorized forms for the same 
rotation operator 
\begin{eqnarray}
\exp (-i\theta \mathbf{nJ}) &\equiv&U= \label{ORDER} \\
&=&\exp (i\theta n_{-}J_{+}-i\theta n_{0}J_{0}+i\theta n_{+}J_{-}) \nonumber \\
&=&\exp (i\alpha _{-}J_{+})\exp (-i\alpha _{0}J_{0})\exp (i\alpha _{+}J_{-}) \nonumber \\
&=&\exp (i\beta _{+}J_{-})\exp (-i\beta _{0}J_{0})\exp (i\beta _{-}J_{+}). \nonumber
\end{eqnarray}
The indices show the  components of coordinates in the cyclic basis: 
\begin{equation}
\mathbf{nJ} = -n_{-}J_{+}+n_{0}J_{0}-n_{+}J_{-},
\label{scalarproduct}
\end{equation}
where
\begin{eqnarray}
J_{+} &=&\frac{-1}{\sqrt{2}}(J_{1}+iJ_{2}), \nonumber\\
J_{-} &=&\frac{1}{\sqrt{2}}(J_{1}-iJ_{2}), \nonumber\\
J_{0} &=&J_{3}. \label{sphericalcoordinates}
\end{eqnarray}
Throughout the present paper the used notations are consistent with those of Ref.~\cite{VMH}.
The spherical components of the unit vector are: 
\begin{eqnarray}
n_{\pm } &=&\mp \frac{1}{\sqrt{2}}\sin (\Theta )\exp (\pm i\Phi ),\nonumber \\
n_{0} &=&\cos (\Theta ), \label{unit}
\end{eqnarray}
where $\Theta $ and $\Phi $ are the polar and azimuthal angles. The unit
length of the vector $\mathbf{n}$ implies, in the cyclic basis, 
\begin{equation}
-2n_{-}n_{+}+n_{0}^{2}=1.
\label{NORMA} 
\end{equation}
The decomposition (\ref{ORDER}) resembles the representation of rotation as
a product of three rotations described by the Euler angles. The
difference is that here all three rotation generators are involved, and they
are non-Hermitian operators.

The relation between the parameters $\theta ,\mathbf{n}$ and $\alpha
_{\mu }$ can be found by taking the derivative over the $\theta $ of the
left and right sides in Eq. (\ref{ORDER}) and using the commutation
relations for the rotation generators. The derivative over $\theta $
gives 
\begin{eqnarray*}
&&(in_{-}J_{+}-in_{0}J_{0}+in_{+}J_{-})\exp (-i\theta \mathbf{nJ}) = \\
&\;&\;\;\;\;\;\;\;\;\;\;\;\;\;i\alpha_{-}^{\prime }J_{+}\exp (i\alpha _{-}J_{+})\exp (-i\alpha _{0}J_{0})\exp(i\alpha _{+}J_{-})\\
&\;&\;\;\;\;\;\;\;\;\;\;\;-i\alpha _{0}^{\prime }\exp (i\alpha _{-}J_{+})J_{0}\exp (-i\alpha
_{0}J_{0})\exp (i\alpha _{+}J_{-})\\
&\;&\;\;\;\;\;\;\;\;\;\;\;+i\alpha _{+}^{\prime }\exp (i\alpha _{-}J_{+})\exp (-i\alpha_{0}J_{0})J_{-}\exp (i\alpha _{+}J_{-}).
\end{eqnarray*}
Here the symbol "$^{\prime }$" stands for the derivative with respect to the variable  
$\theta $. Using the commutation relations satisfied by the angular momentum components~\cite{VMH} 
\begin{eqnarray}
\lbrack J_{+},J_{-}] &=&-J_{0},  \nonumber \\
\lbrack J_{\pm},J_{0}] &=&\mp J_{\pm}.  
  \label{COMMUTANGULAR}
\end{eqnarray}
one finds 
\begin{eqnarray}
\exp (i\alpha _{-}J_{+})J_{0}\exp (-i\alpha _{-}J_{+}) &=&
J_{0}-i\alpha _{-}J_{+},  \label{CR1} \\
\exp (-i\alpha _{0}J_{0})J_{-}\exp (i\alpha _{0}J_{0}) &=& 
J_{-}\exp(i\alpha _{0}),  \nonumber\\
\exp (i\alpha _{-}J_{+})J_{-}\exp (-i\alpha _{-}J_{+}) &=& 
J_{-}-i\alpha _{-}J_{0}-\frac{\alpha _{-}^{2}}{2}J_{+}. \nonumber
\end{eqnarray}
In deriving the above relations we made use of the following equation:
\[
e^ABe^{-A}=\sum_{k=0}^{\infty}\frac{1}{k!}
\underbrace{[A,[A,...[A}_{k},B]...]].
\]
which holds for any operators $A$  and $B$.
One thus arrives at the equation 
\begin{widetext}
\begin{eqnarray}
in_{-}J_{+}-in_{0}J_{0}+in_{+}J_{-}=i\alpha _{-}^{\prime }J_{+}-i\alpha_{0}^{\prime }(J_{0}-i\alpha _{-}J_{+})
+i\alpha _{+}^{\prime }\exp (i\alpha_{0})(J_{-}-i\alpha _{-}J_{0}-\frac{\alpha _{-}^{2}}{2}J_{+}), 
\end{eqnarray}
\end{widetext}
that can be split into three equations
\begin{eqnarray}
n_{-} &=&\alpha _{-}^{\prime }+i\alpha _{0}^{\prime }\alpha _{-}-\alpha
_{+}^{\prime }\exp (i\alpha _{0})\frac{\alpha _{-}^{2}}{2},  \nonumber \\
n_{0} &=&\alpha _{0}^{\prime }+i\alpha _{+}^{\prime }\exp (i\alpha
_{0})\alpha _{-},   \nonumber \\
n_{+} &=&\alpha _{+}^{\prime }\exp (i\alpha _{0}), \label{ODE}
\end{eqnarray}
which can further  be  brought to the equivalent form 
\begin{eqnarray}
n_{-} &=&\alpha _{-}^{\prime }+in_{0}\alpha _{-}+\frac{n_{+}\alpha _{-}^{2}}{2}, \nonumber\\
n_{0} &=&\alpha _{0}^{\prime }+in_{+}\alpha _{-},\nonumber \\
n_{+} &=&\alpha _{+}^{\prime }\exp (i\alpha _{0}).
\end{eqnarray}
Using the boundary conditions 
\[
\alpha _{-}(0)=\alpha _{0}(0)=\alpha _{+}(0)=0, 
\]
we obtain 
\begin{eqnarray}
\alpha _{\pm }(\theta ) &=&\frac{n_{\pm }2\tan (\theta /2)}{1+in_{0}\tan
(\theta /2)},  \label{ALPHAPM} \\
\alpha _{0}(\theta ) &=&-2i\ln (\cos (\theta /2)+in_{0}\sin (\theta /2)).
\label{ALPHAZE}
\end{eqnarray}

The result for the functions $\alpha _{\mu }(\theta )$, depends only on the commutation relations 
of $J_{\mu }$ and therefore is the same for any representation of the   $SU(2)$ 
algebra generated by the angular momentum  operators. Let us make use
of this remark. Consider the fundamental representation and replace the operators 
$J_{\mu }$ by the Pauli matrices divided by 2. Next, we expand the left and
right sides of the expression (\ref{ORDER}) in the parameters $\theta $ and 
$\alpha _{\mu }$. The decomposition over $\alpha _{\pm }$ is necessarily truncated at the
terms linear in $\alpha _{\pm }$ due the equation $J_{\pm }^{2}=0$  valid for the spin-1/2 representation.
Thus, one obtains a system of algebraic equations for $\alpha _{\mu }(\theta )$
whose solutions as can easily be checked are given by equations 
(\ref{ALPHAPM}) and (\ref{ALPHAZE}). 
The algebraic method is fully equivalent to solving the system of ordinary differential equations (\ref{ODE}).

We assumed, so far, that the rotation angle $\theta $ and the unit vector $\mathbf{n}$ 
are real quantities. However, in the above development the  condition for the mentioned variables to be real
was not explicitly used. Equations (\ref{ALPHAPM}) and (\ref{ALPHAZE}), therefore, can
be analytically continued to complex values of the parameter $\theta $, the
vectors $\mathbf{n}\,$ could also be complex under the condition that their
square is equal to unity, i.e., $\mathbf{n}^{2}=-2n_{+}n_{-}+n_{0}^{2}=1$. Obviously,
these  are the most general conditions. Any operator that can be
written as the exponential of a linear combination of generators of the
rotation with complex coefficients can be represented using a complex
parameter $\theta $ and a unit complex vector $\mathbf{n}$.

In calculating the state norms and the matrix elements of operators 
one needs to know another factorized form for the rotation operator 
where the factor operators show up in a reverse order as compared with that 
given in the second line of Eq.~(\ref{ORDER}). The reverse order is shown
explicitly in the third line of Eq.~(\ref{ORDER}). We thus are looking for the 
 rotation parameters $\beta_{\mu}$ as functions of $\theta$ and $\mathbf{n}$. 

In the derivation of equations (\ref{ALPHAPM}) and (\ref{ALPHAZE}) we have
used only commutation relations for the generators of rotations. The
explicit form of $\beta _{\mu }(\theta )$ can be obtained by using the fact
that under the replacement of $(J_{+},J_{-},J_{0})\rightarrow
(J_{-},J_{+},-J_{0})$ the commutation relations  (\ref{COMMUTANGULAR}) remain unchanged. From this
we immediately find
\begin{eqnarray}
\beta _{\pm }(\theta ) &=&\frac{n_{\pm }2\tan (\theta /2)}{1-in_{0}\tan
(\theta /2)},  \label{BETAPM} \\
\beta _{0}(\theta ) &=&2i\ln (\cos (\theta /2)-in_{0}\sin (\theta /2)).
\label{BETAZE}
\end{eqnarray}

If we are able to express $\beta _{\mu }$ through the $\alpha _{\mu }$, then
we can change the order of the exponents with different operators $J_{\mu }$. 
These relations are, however, simple:
\begin{eqnarray}
\beta _{\pm } &=&-\alpha _{\mp }^{*},  \label{ALBEPM} \\
\beta _{0} &=&\alpha _{0}^{*}.  \label{ALBEZE}
\end{eqnarray}

Equations (\ref{ALBEPM}) and (\ref{ALBEZE}) are the necessary and sufficient in order that the operator $U$ is unitary.
 They are  therefore valid for real $\theta $ and
for real unit vectors $\mathbf{n}$, that is for pure rotation. These
equations cannot be continued analytically to the complex values of $\theta $
and $\mathbf{n}$, because they involve the operation of complex conjugation.
In the case of complex rotations, one should return to equations (\ref
{ALPHAPM}) and (\ref{ALPHAZE}). We express the complex parameters $\theta $
and $\mathbf{n}$ through $\alpha _{\mu }$, and substitute these expressions
in formulas (\ref{BETAPM}) and (\ref{BETAZE}). Simple calculations give 
\begin{eqnarray}
\beta _{\pm } &=&\frac{\alpha _{\pm }\exp (i\alpha _{0})}{1+\frac{1}{2}%
\alpha _{+}\alpha _{-}\exp (i\alpha _{0})},  \label{ALPHA2BETAPM} \\
\beta _{0} &=&\alpha _{0}+2i\ln (1+\frac{1}{2}\alpha _{+}\alpha _{-}\exp
(i\alpha _{0})).  \label{ALPHA2BETAZE}
\end{eqnarray}
These formulas establish the connection between arbitrary complex parameters 
$\alpha _{\mu }$ and $\beta _{\mu }$. The inverse relations have the form 
\begin{eqnarray}
\alpha _{\pm } &=&\frac{\beta _{\pm }\exp (-i\beta _{0})}{1+\frac{1}{2}\beta
_{+}\beta _{-}\exp (-i\beta _{0})},  \label{BETA2ALPHAPM} \\
\alpha _{0} &=&\beta _{0}-2i\ln (1+\frac{1}{2}\beta _{+}\beta _{-}\exp
(-i\beta _{0})).  \label{BETA2ALPHAZE}
\end{eqnarray}

The factorized expression of any rotation like operator is known in the literature as 
the Baker-Cambell-Haussdorff formula. The general necessary conditions which make this factorization possible are discussed in Ref.~\cite{Kirz}.

 Inverse relations allowing de-factorization of the rotation look as follows:
\begin{eqnarray}
\cos (\theta /2)&=&\cos (\alpha_{0}/2)+\frac{1}{4}\alpha_{+}\alpha_{-}\exp (  i\alpha_{0}/2) \nonumber \\
                &=&\cos (\beta _{0}/2)+\frac{1}{4}\beta _{+}\beta _{-}\exp (- i\beta _{0}/2), \nonumber \\
n_{\pm}\sin (\theta /2)&=&\frac{1}{2}\alpha_{\pm}\exp (  i\alpha_{0}/2) \nonumber \\
       &=&\frac{1}{2}\beta _{\pm}\exp (- i\beta _{0}/2), \\
n_{0}\sin (\theta /2)&=&\sin (\alpha _{0}/2)+\frac{i}{4}\alpha _{+}\alpha _{-}\exp (i\alpha _{0}/2)  \nonumber \\
                     &=&\sin (\beta  _{0}/2)+\frac{i}{4}\beta  _{+}\beta  _{-}\exp (-i\beta  _{0}/2). \nonumber
\end{eqnarray}

\renewcommand{\theequation}{3.\arabic{equation}}
\setcounter{equation}{0}
\label{sec:level3}
\section{BCS state}

The BCS wave function can
be written as a unitary transformation of the vacuum state. Such a form
significantly simplifies the calculations and gives physical meaning to
the algebraic transformations involved in  the formalism. 

In the analogy with Eq.~(\ref{rotoper}) one can write
\begin{equation}
|BCS\rangle =e^{-iF}|0\rangle , 
\end{equation}
where 
\begin{equation}
F= \frac{1}{\sqrt{2}}\sum_{\alpha }\left(   z_{\alpha}  \sum_{m}c_{\alpha m}^{\dagger }d_{\alpha \tilde{m}}^{\dagger }
                                         + z_{\alpha}^*\sum_{m}d_{\alpha \tilde{m}}c_{\alpha m}\right)  \label{F}
\end{equation}
is a Hermitian operator while $|0\rangle$ denotes the bare vacuum state. Here, $c_{\alpha m}^{\dagger }$ and $d_{\alpha
m}^{\dagger }$ are particle creation operators for protons and neutrons respectively, $\alpha $ is the index
numbering shells. 
For example, in a spherical shell model $ \alpha $ is the set of quantum numbers $(njl) $, where $n$ is the radial quantum number, 
$j$ is the total angular momentum, and $l$ is the orbital angular momentum. 
$\tilde{m}$ denotes the time reversal sub-state: $d_{\alpha \tilde{m}}^{\dagger }=(-)^{j - m}d_{\alpha, -m}^{\dagger }$.
For alike nucleons in Ref.~\cite{Rad84} and for a generalized $pn$ pairing interaction in Ref.~\cite{RaMo} analogous, otherwise different, unitary transformations have been used.

The  creation operators images through the transformation  $\exp(-iF)$ define the quasiparticle creation operators:

\begin{eqnarray}
e^{-iF}c_{\alpha m}^{\dagger }e^{iF} &=&\mathfrak{c}_{\alpha m}^{\dagger },\nonumber \\
e^{-iF}d_{\alpha m}^{\dagger }e^{iF} &=&\mathfrak{d}_{\alpha m}^{\dagger }. \label{BV}
\end{eqnarray}
with:
\begin{eqnarray}
\mathfrak{c}_{\alpha m}^{\dagger }=     \cos (\frac{|z_{\alpha }|}{\sqrt{2}})c_{\alpha m}^{\dagger } 
- i\frac{z_{\alpha }^*}{|z_{\alpha }|}\sin (\frac{|z_{\alpha }|}{\sqrt{2}})d_{\alpha \tilde{m}},\nonumber\\
\mathfrak{d}_{\alpha m}^{\dagger }     =\cos (\frac{|z_{\alpha }|}{\sqrt{2}})d_{\alpha m}^{\dagger }
- i\frac{z_{\alpha }^*}{|z_{\alpha }|}\sin (\frac{|z_{\alpha }|}{\sqrt{2}})c_{\alpha \tilde{m}}.
\end{eqnarray}
We recognize here the Bogoliubov-Valatin (BV) transformation:
\begin{eqnarray}
\mathfrak{c}_{\alpha m}^{\dagger }&=&U^{p}_{\alpha}c_{\alpha m}^{\dagger }-V^{p}_{\alpha}d_{\alpha \tilde{m}},
\nonumber\\
\mathfrak{d}_{\alpha m}^{\dagger }&=&U^{n}_{\alpha}d_{\alpha m}^{\dagger }-V^{n}_{\alpha}c_{\alpha \tilde{m}}.
\end{eqnarray}
Using the polar representation of the complex variable $z$,
\begin{equation}
z_{\alpha }= \frac{1}{\sqrt{2}}\rho _{\alpha }e^{-i\psi _{\alpha }},
\label{azim}
\end{equation}
the occupation and non-occupation probability coefficients, $V$ and $U$ respectively, 
are expressed as follows:
\begin{eqnarray}
U^{\tau}_{\alpha}&=&\cos (\frac{\rho_{\alpha}}{2}),\nonumber\\
V^{\tau}_{\alpha}&=&    e^{i(\psi_{\alpha} + \frac{\pi}{2})}  \sin (\frac{\rho_{\alpha}}{2}), \;\tau =p,n, 
\label{UandV}
\end{eqnarray}
The complex variable $z_{\alpha}$ can be interpreted as the $\mu = -1$ 
component of a representative vector from the associated classical phase space.
From Eq.~(\ref{UandV}) it comes out that the BV transformation coefficients satisfy the normalization conditions:
\begin{equation}
U_{\alpha}^2+|V_{\alpha}|^2 =1,
\end{equation}
which assure that the quasiparticle operators  obey anticommutation relations specific to fermions.
Since $U^p_{\alpha}=U^n_{\alpha}$ and $V^p_{\alpha}=V^n_{\alpha}$, in what follows we shall omit 
the isospin index for the BV transformation coefficients.
Applying the operator  $\exp(-iF)$ on the obvious equations $c_{\alpha m}|0\rangle =d_{\alpha m}|0\rangle =0$ one obtains $\mathfrak{c}_{\alpha m}|BCS\rangle =\mathfrak{d}_{\alpha m}|BCS\rangle
=0, $ which expresses the fact that the BCS state is a vacuum state for the quasiparticle operators. 

Using the proton and neutron creation operators one could define the bi-linear operators 
\begin{eqnarray}
J_{\alpha  +} &=&- \frac{1}{\sqrt{2}}\sum_{m}c_{\alpha m}^{\dagger }d_{\alpha \tilde{m}}^{\dagger }, \\
J_{\alpha  - } &=&- \frac{1}{\sqrt{2}}\sum_{m}c_{\alpha m}d_{\alpha \tilde{m}},
\\
J_{ \alpha  0} &=&\frac{1}{2}\sum_{m}\left( c_{\alpha m}^{\dagger }c_{\alpha
m}-d_{\alpha m}d_{\alpha m}^{\dagger }\right) ,
\end{eqnarray}
which satisfy the $SU(2)$ algebra commutation relations (cf. Eqs. (\ref{COMMUTANGULAR})) :
\begin{eqnarray}
\lbrack J_{\alpha  +},J_{ \beta  -}] &=&-\delta _{\alpha \beta }J_{\alpha  0}, \nonumber\\
\lbrack J_{\alpha  \pm},J_{\beta 0}] &=&\mp\delta _{\alpha \beta }J_{\alpha  \pm}. 
\label{quasispin}
\end{eqnarray}
We conventionally call these operators as quasispin operators. Indeed if we replace the $d$
operators by $c$ the resulting algebra defines the proton senioritiy states. 
Moreover replacing the $c$ operators by $d$ one obtains the neutron quasipin algebra 
which defines the neutron seniority scheme.  Although there is a danger of mixing them up 
with angular momentum operator we use the notation $J$ for the quasispin algebra operators. 
Actually the $pn$ pairing operators and the total number of nucleons form a representation 
of the $SU(2)$ algebra which is different from that
generated by angular momentum components. Due to Eq. (\ref{quasispin}),the transformation $e^{-iF}$ acquires the significance of a quasirotation. Within this context the quasiparticle operators
appear to be the result of a quasirotation applied to the creation and annihilation particle operators.

It is worth mentioning some useful properties: $J{}_{\alpha \pm }^{\dagger }=-J_{\alpha \mp }$, $J_{\alpha 0}$
is a Hermitian operator. Due to the Pauli principle both $J_{\alpha + }$ and $J_{\alpha - }$ are nilpotent: 
\[
\left( J_{\alpha  - }\right) ^{2j_{\alpha }+2}=\left( J_{\alpha  +}\right)^{2j_{\alpha }+2}=0. 
\]
Acting on the vacuum state, one has
\begin{eqnarray*}
J_{\alpha  - }|0\rangle &=&0, \\
J_{\alpha 0}|0\rangle &=&-(j_{\alpha }+\frac{1}{2})|0\rangle .
\end{eqnarray*}
For a given $\alpha$ the eigenstates of the proton-neutron pairing Hamiltonian in the restricted single particle space can be expressed in terms of the irreducible representations of the $SU(2)$ group. These are the states $|J_{\alpha},J_{\alpha 0}\rangle$ which are simultaneous eigenstates for the operators $J^2_{\alpha}$ and $J_{\alpha 0}$. The interpretation of the quasispin comes out
from the following obvious equations:
\begin{eqnarray}
J_{\alpha -}|J_{\alpha}, -J_{\alpha}\rangle &=&0,\nonumber\\
J_{\alpha 0}|J_{\alpha}, J_{\alpha 0}\rangle &=&J_{\alpha 0}|J_{\alpha}, J_{\alpha 0}\rangle
\nonumber\\
&=&\left(\frac{N}{2}-\frac{1}{2}(2j_a+1)\right)|J_{\alpha}, J_{\alpha 0}\rangle,\nonumber\\
J_{\alpha 0}&=&\frac{N}{2} -\frac{1}{2}(2j_{\alpha}+1).
\end{eqnarray}
Since the values for $N$ range from 0 to $2(2j_{\alpha}+1)$ it results that the minimum and maximum values of $J_{\alpha 0}$ are $-(2j_{\alpha} +1)/2$ and $(2j_{\alpha} +1)/2$, respectively. Consequently, the quasispin has the expression:
\begin{equation}
J_{\alpha}=\frac{1}{2}(2j_{\alpha}+1).
\end{equation}
Thus, the state with a minimum quasispin projection to the $z$ axis is a kind of Hartree-Fock vacuum for the lowering quasispin operator. Also, the component $z$ of quasispin is related to the state angular momentum and the total number of particles, $N$, distributed on the substates of the single shell,$j_{\alpha}$, while the quasispin is given by the semi-degeneracy of the given single particle state.

The transformation $e^{-iF}$ is very useful for calculating operator matrix elements either in the quasiparticle or in the particle representation.
As an example, we find  the matrix element  of a unit operator between the states 
$\exp(i\beta _{\alpha -}J_{\alpha  +})|0\rangle $ and $\langle 0|\exp (i\beta_{\alpha +}J_{\alpha  - })$.

By means of Eqs.~(\ref{BETA2ALPHAPM}) and (\ref{BETA2ALPHAZE}), we obtain 
\begin{eqnarray}
&&\hspace{-6mm}\langle 0|\exp (i\beta _{\alpha +}J_{\alpha  - })\exp (i\beta _{\alpha -}J_{\alpha  +})|0\rangle = \nonumber \\
&=&\langle 0|\exp (i\alpha _{\alpha -}J_{\alpha  +})\exp (-i\alpha _{\alpha 0}J_{\alpha 0})\exp (i\alpha _{\alpha +}J_{\alpha  - })|0\rangle \nonumber  \\
&=&\langle 0|\exp (-i\alpha _{\alpha 0}J_{\alpha 0})|0\rangle = \langle 0|\exp (i\alpha _{\alpha 0}(j_{\alpha }+\frac{1}{2}))|0\rangle \nonumber \\
&=& \left( 1+\frac{1}{2}\beta _{\alpha +}\beta _{\alpha -}\right)^{2j_{\alpha }+1}. 
\label{mostimportantaverage}
\end{eqnarray}

In the above case, the parameter $\beta _{\alpha 0}$ is equal to zero; we
used Eq. (\ref{BETA2ALPHAZE}) to express $\alpha _{\alpha 0}$ in terms of $%
\beta _{\alpha +}\ $and $\beta _{\alpha -}$.

Under the unitary transformation, the generators $J_{\mu \alpha}$ become: 
\[
e^{-iF}J_{\alpha \mu }e^{iF}=\mathfrak{J}_{\alpha \mu}, \;\mu=\pm,0,
\]
where
\begin{eqnarray}
\mathfrak{J}_{ \alpha +} &=&- \frac{1}{\sqrt{2}}\sum_{m}\mathfrak{c}_{\alpha m}^{\dagger }\mathfrak{d}_{\alpha \tilde{m}}^{\dagger },\nonumber \\
\mathfrak{J}_{ \alpha -} &=&- \frac{1}{\sqrt{2}}\sum_{m}\mathfrak{c}_{\alpha m}\mathfrak{d}%
_{\alpha \tilde{m}}, \nonumber\\
\mathfrak{J}_{\alpha 0} &=&\frac{1}{2}\sum_{m}\left( \mathfrak{c}_{\alpha
m}^{\dagger }\mathfrak{c}_{\alpha m}-\mathfrak{d}_{\alpha m}\mathfrak{d}_{\alpha
m}^{\dagger }\right) ,
\end{eqnarray}

Under the action of a unitary BV transformation, the operator $\exp(-iF)$ maps onto itself, since
\begin{eqnarray}
\mathfrak{F}&\equiv& e^{-iF}F e^{iF} \nonumber \\
        &=& \frac{1}{\sqrt{2}}\sum_{\alpha}\left(z_{\alpha}  \sum_{m}\mathfrak{c}_{\alpha
m}^{\dagger }\mathfrak{d}_{\alpha \tilde{m}}^{\dagger }+z_{\alpha}^*\sum_{m}\mathfrak{d}_{\alpha \tilde{m}}\mathfrak{c}_{\alpha m}\right) \nonumber \\
        &=& F.
\end{eqnarray}
 The reciprocal relation for Eq.~(\ref{BV}) takes the form
\begin{eqnarray}
c^{\dagger}_{\alpha m}&=&e^{i\mathfrak{F}}\mathfrak{c}^{\dagger}_{\alpha m}e^{-i\mathfrak{F}},\nonumber\\
d^{\dagger}_{\alpha m}&=&e^{i\mathfrak{F}}\mathfrak{d}^{\dagger}_{\alpha m}e^{-i\mathfrak{F}}
\end{eqnarray}
or, explicitly,
\begin{eqnarray}
c^{\dagger}_{\alpha m}=U_{\alpha}\mathfrak{c}^{\dagger}_{\alpha m}+V_{\alpha}\mathfrak{d}_{\alpha \tilde{m}},\nonumber\\
d^{\dagger}_{\alpha m}=U_{\alpha}\mathfrak{d}^{\dagger}_{\alpha m}+V_{\alpha}\mathfrak{c}_{\alpha \tilde{m}}.
\end{eqnarray}

\renewcommand{\theequation}{4.\arabic{equation}}
\setcounter{equation}{0}
\label{sec:level4}
\section{Particle number projection}

The projection operator to the state with a definite number of particles is
given by
\begin{equation}
P_{N}=\int\limits_{0}^{2\pi} \frac{d\varphi }{2\pi }e^{i(\hat{N}-N)\varphi }. 
\label{proj}
\end{equation}
One can check that the signature property of a  projection operator is satisfied: 
\begin{equation}
P_{N}P_{N}=P_{N}.  \label{PP=P}
\end{equation}
The particle number operator can be expressed in terms of the operators 
$J_{ \alpha  0}$ 
\begin{equation*}
\hat{N}=\sum_{\alpha }\hat{N}_{\alpha },
\end{equation*}
where
\begin{equation}
\hat{N}_{\alpha }=2J_{ \alpha  0}+2j_{\alpha }+1.  
\label{NUMOPE}
\end{equation}

The operator $P_{N}$ acting on the  BCS wave function gives a
state with a definite number of particles 
\begin{eqnarray}
|BCS,N\rangle &\equiv& C_{N}P_{N}|BCS\rangle \nonumber \\
&=&C_{N} \int\limits_{0}^{2\pi} \frac{d\varphi }{2\pi }e^{i(\hat{N}-N)\varphi }e^{-iF}|0\rangle \nonumber\\
&=&C_{N}\int\limits_{0}^{2\pi} \frac{d\varphi }{2\pi }e^{-iN\varphi }e^{-iF(\varphi )}|0\rangle , \label{def}
\end{eqnarray}
where 
\begin{equation}
F(\varphi ) = -\sum_{\alpha }\left( z_{\alpha }e^{2i\varphi }J_{\alpha  +} - z_{\alpha }^*e^{-2i\varphi }J_{\alpha  - }\right)
\end{equation}
and $F=F(0).$

Using the representation (\ref{azim}), the operator $F(\varphi )$ 
can be written as a sum of scalar products of vectors $\rho _{\alpha } \mathbf{n}_{\alpha}$ and quasispin operators $\mathbf{J}_{\alpha}$:
\begin{eqnarray}
F(\varphi )= -\sum_{\alpha }\rho _{\alpha }\left(n_{\alpha -}J_{\alpha  +} +n_{\alpha +}J_{\alpha  - }\right). 
\end{eqnarray}
Here, $\mathbf{n}_{\alpha}$ are unit vectors 
defined by the cyclic coordinates
\begin{eqnarray}
n_{\alpha \pm } &=&\mp \frac{1}{\sqrt{2}}\exp (\pm i(\psi _{\alpha}- 2\varphi )), \nonumber \\
n_{\alpha 0} &=&0. \label{asis}
\end{eqnarray}

Using Eq. (\ref{ORDER}), we obtain 
\begin{eqnarray}
\exp (-iF(\varphi )) &=& \exp (i\sum_{\alpha }\rho _{\alpha }\left( n_{\alpha-}J_{\alpha  +}+n_{\alpha +}J_{\alpha  - }\right) ) \nonumber \\
                     &=& \exp (i\sum_{\alpha }\alpha _{\alpha -}J_{\alpha  +}) \nonumber \\
                     &\times& \exp (-i\sum_{\alpha}\alpha _{\alpha 0}J_{\alpha 0}) \exp (i\sum_{\alpha }\alpha _{\alpha +}J_{\alpha  - }), \nonumber 
\end{eqnarray}
where, according to Eqs. (\ref{ALPHAPM}) and (\ref{ALPHAZE}), 
\begin{eqnarray}
\alpha _{\alpha \pm } &=&n_{\alpha \pm }2\tan (\rho _{\alpha }/2),\nonumber \\
\alpha _{\alpha 0} &=&-2i\ln (\cos (\rho _{\alpha }/2)).
\end{eqnarray}

Acting with the operator $\exp (-iF(\varphi ))$, after the factorizing it, on the vacuum state, we obtain
\begin{eqnarray*}
\exp (-iF(\varphi ))|0\rangle &=& \prod_{\alpha }\left( \cos ^{2}(\rho _{\alpha }/2)\right) ^{j_{\alpha}+1/2} \\
&\times& \exp ( i \sum_{\beta} n_{\beta -}2\tan (\rho _{\beta }/2)J_{\beta +})|0\rangle . \nonumber 
\end{eqnarray*}

Now, we are in a position to find the projected state with $N$ nucleons. 
\begin{widetext}
\begin{eqnarray}
|BCS,N\rangle &=& C_{N} \int\limits_{0}^{2\pi }\frac{d\varphi }{2\pi }\exp (-iN\varphi )
\prod_{\alpha}\left( \cos ^{2}(\rho _{\alpha }/2)\right) ^{j_{\alpha }+1/2} 
\exp(i\sum_{\alpha }\exp (i(-\psi _{\alpha }+2\varphi ))\sqrt{2}\tan (\rho_{\alpha }/2)J_{\alpha +})|0\rangle  \label{BCSNstate}\nonumber\\
&=&C_{N}\prod_{\alpha }\left( \cos ^{2}(\rho _{\alpha }/2)\right)
^{j_{\alpha }+1/2}\int\limits_{C}^{}\frac{d\zeta}{2\pi i}\frac{1}{\zeta^{N+1}}
\exp (i\zeta^{2}\sum_{\alpha }\exp (-i\psi _{\alpha })\sqrt{2}\tan (\rho
_{\alpha }/2)J_{\alpha +})|0\rangle  \nonumber \\
&=&C_{N}\prod_{\alpha }\left( \cos ^{2}(\rho _{\alpha }/2)\right)
^{j_{\alpha }+1/2}\frac{1}{\left( N/2\right) !}
\left( i\sum_{\alpha }\exp(-i\psi _{\alpha })\sqrt{2}\tan (\rho _{\alpha }/2)J_{\alpha +}\right)
^{N/2}|0\rangle .  \nonumber
\end{eqnarray}
\end{widetext}
The contour $C$ encompasses the point $\zeta = 0$, the integration is performed in the direction of increasing $\varphi = \arg \zeta$.

By the condition (\ref{PP=P}), the problem of finding the normalization constant reduces to  calculating 
the overlap of unprojected and projected states: 
\begin{widetext}
\begin{eqnarray}
\langle BCS,N|BCS,N\rangle &=& C_{N}\langle BCS|BCS,N\rangle 
= C_{N}^{2}\prod_{\alpha } \left( \cos ^{2}(\rho _{\alpha }/2)\right) ^{2j_{\alpha }+1} \int\limits_{C}^{}\frac{d\zeta}{2\pi i} \frac{1}{\zeta^{N+1}}\nonumber\\
&\times&\langle 0|\exp ( i \sum_{\alpha} \exp (i \psi _{\alpha})\sqrt{2}
\tan (\rho _{\alpha}/2)J_{\alpha -}) 
\exp (i\zeta^{2}\sum_{\alpha }\exp (-i \psi _{\alpha })\sqrt{2}\tan (\rho _{\alpha }/2)
J_{\alpha +}) |0\rangle \nonumber\\
&=& C_{N}^{2}\int\limits_{C}^{}\frac{d\zeta}{2\pi i}\frac{1}{\zeta^{N+1}}
\prod_{\alpha }\left( \cos ^{2}(\rho _{\alpha }/2)+\zeta^{2}\sin ^{2}(\rho
_{\alpha }/2)\right) ^{2j_{\alpha }+1}. \label{inter}
\end{eqnarray}
From the above equation we obtain
\begin{equation}
C_N^{-2}=\int\limits_{C}^{}\frac{d\zeta}{2\pi i}\frac{1}{\zeta^{N+1}}
\prod_{\alpha }\left( \cos ^{2}(\rho _{\alpha }/2)+\zeta^{2}\sin ^{2}(\rho
_{\alpha }/2)\right) ^{2j_{\alpha }+1}.
\label{intQ}
\end{equation}
\end{widetext}

In the derivation of the expression (\ref{inter}), we used the Eqs.~(\ref{BETA2ALPHAPM}) and 
(\ref{BETA2ALPHAZE}), 
which allow to change the order of the factors of the unitary operator.
Also, we had to calculate the average of the product of exponents of
operators associated to individual shells. If the two involved shells  are distinct the
exponents commute with each other and their average over the vacuum is unity. An average value
different from unity occurs only when $\alpha ^{\prime }=\alpha^{\prime \prime} $, i.e. only when the two
shells coincide. The average value splits thereby into the product of the
average values for the individual shells.

Further simplification can be achieved by using a binomial
formula and then evaluating the integrand residue in the point $\zeta=0$.
After applying the binomial formula for the sum of a large number of terms
and then finding the residue of the integrand the problem becomes
combinatorial in nature, which is not attractive from the computational point
of view, because the number of options needed to be considered and the
number of terms in the sum grow with $N$ exponentially. 

From the computational point of view, the possibility of reducing the problem to evaluation of
a recursion looks more attractive. We introduce the  function 
\begin{equation}
Q(N)=C_{N}^{-2}
\end{equation}
with $C_{N}^{-2}$ defined above.

Integrating by parts, one finds 
\begin{equation}
Q(N)= \sum_{\beta } \int\limits_{C}\frac{d\zeta}{2\pi i}\frac{\mathcal{G}_{\beta}(\zeta)}{\zeta^{N+1}},
\label{IQN}
\end{equation}
where
\begin{eqnarray}
\hspace{-9mm}\mathcal{G}_{\beta}(\zeta) &=& \frac{\Omega_{\beta}}{N} \frac{\zeta^{2}\sin ^{2}(\rho _{\beta }/2)}{\cos
^{2}(\rho _{\beta }/2)+\zeta^{2}\sin ^{2}(\rho _{\beta }/2)} \nonumber \\
\hspace{-9mm}&\times& \prod_{\alpha}\left( \cos ^{2}(\rho _{\alpha }/2)+\zeta^{2}\sin ^{2}(\rho _{\alpha }/2)\right) ^{2j_{\alpha }+1},
\label{QN}
\end{eqnarray}
with $\Omega_{\beta} = 2(2j_{\beta} + 1)$. 
We expand further the expression in front of the product sign in powers of 
$\zeta^{2}$. Each member of the series is a function of $Q(N^{\prime })$ for
some value of $N^{\prime }<N$. The function $Q(N)$, therefore, is expressed
as the sum of $Q(N^{\prime })$ evaluated for a smaller number of particles.
It only remains to fix the boundary value for $N=0$. From the definition of $Q(N)$ 
one easily finds 
\begin{equation}
Q(0)=\prod_{\alpha }(\cos ^{2}(\rho _{\alpha }/2))^{2j_{\alpha }+1}. 
\end{equation}
Note that for  negative integer values of $N$, the function $Q(N)$ is equal to zero, as can be
seen from the Cauchy theorem related to the contour integral. Also, 
$Q(N)=0 $ for $N=1$ mod$(2)$.  We thus get a recursion
\hspace{-6mm}
\begin{eqnarray}
Q(N)&=& \sum_{\beta }Q^{\beta }(N),                                                                          \label{QNrecursion0} \\
Q^{\beta }(N) &=& \frac{\Omega_{\beta}}{N} \sum_{n=1}^{N/2}(-)^{n+1}\tan^{2n}(\rho _{\beta }/2)Q(N-2n). \nonumber \\
\label{QNrecursion1}
\end{eqnarray}
The number of operations to calculate $Q(N)$ grows with increasing $N$ only
quadratically. From the viewpoint of numerical calculation, estimates for
oscillatory contour integrals are associated with considerable difficulties. We
avoid this difficulty by reducing the problem to the computation 
 of the recursion relations. Within a variation after projection procedure 
the angles $\rho_{\beta}$ are determined by the equations provided by the conditions 
that the energy of the system for a fixed $N$ be minimum. 
The same angles are however used to calculate the factors $Q(N-2n-2)$ involved in the summation 
operation over $n$. The BCS angles as well as the number of involved shells are preserved during the iteration process. 

The product factors from Eq. (\ref{QN}) can be expanded in power series of $1/\zeta$. 
In this case we obtain a recursion for calculating the function $Q(N)$ starting from large numbers of particles. Recursion of this form is more convenient 
to calculate $Q(N)$ for values of $N$ close to the maximum
\begin{equation}
\Omega = \sum_{\beta}\Omega_{\beta}.
\end{equation}
Using the the expansion in $1/ \zeta$, we obtain Eq.~(\ref{QNrecursion0}) with
\begin{eqnarray}
Q^{\beta}(N) &=& \frac{\Omega_{\beta}}{\Omega - N}               \label{QNrecursion3} \\
&\times&\sum_{n=1}^{ (\Omega - N)/2 }(-)^{n + 1}\cot ^{2n}(\rho _{\beta }/2)Q(N+2n). \nonumber
\label{QNrecursion3}
\end{eqnarray}
These expressions should be supplemented by the boundary condition
\begin{equation}
Q(\Omega) =\prod_{\alpha }(\sin ^{2}(\rho _{\alpha }/2))^{2j_{\alpha }+1}. 
\label{QNrecursion5}
\end{equation}
For $N> \Omega$, the function $Q(N)$ is identically zero. 

The only singularity of the integrand in the expression  (\ref{IQN})  is a pole at $\zeta = 0$. For this reason, we can deform the contour of integration, 
squeezing it around zero or moving it to infinity.
In the first case, the expansion in powers of $\zeta$ is the appropriate one, while in the second case, the expansion in powers of $1/\zeta$ is valid. 
Obviously, the results (\ref{QNrecursion1}) and (\ref{QNrecursion3}) coincide.
 Also we note that for $N=0$ the function
$Q^{\beta}(N)$ given by Eq. (\ref{QNrecursion1}) is not defined. For this case one should use the
$1/{\zeta}$ expansion from Eq. (\ref{QNrecursion3}). By contrast for calculating 
$Q^{\beta}(\Omega)$ the expression \ref{QNrecursion1} is the appropriate one.
From the integral representation of $Q(N)$ as well as from Eq.\;(A.11) considered for the diagonal
case, a very simple relation follows:
\begin{equation}
\sum_{N=0}^{\Omega}Q(N)=1.
\end{equation}
In virtue of this expression $Q(N)$ acquires the significance of the  admixture probability of the $N$-projected state in the BCS wave function.

The average values of any fermion operator corresponding to  the particle number projected  BCS state,  
can be expressed in terms of the function $Q(N)$, which is the inverse of the normalization constant squared. In next Section this will be shown for few examples.

\renewcommand{\theequation}{5.\arabic{equation}}
\setcounter{equation}{0}
\label{sec:level5}
\section{Average energy}

The projected state might be used as a variational state for a proton-neutron pairing Hamiltonian.
We shall calculate the average of a many body Hamiltonian with two body interaction whit a strength depending on shells
\begin{equation}
\hat{H}=\sum_{\alpha }(\epsilon _{\alpha }-\lambda)N_{\alpha }+\sum_{\alpha \beta }\mathcal{V}_{\alpha \beta }J_{ \alpha  +}J_{\beta -}. 
\label{ModHam}
\end{equation}
Within the standard BCS formalism, $\lambda$ is the Fermi sea level (chemical potential) which is to be fixed by solving the BCS equations. 
Here it is just a parameter involved in the gap energy equation.
The two terms entering the microscopic Hamiltonian will be separately treated. The situation when
 $\mathcal{V}_{\alpha\beta}= -2G$ can be used to describe a possible transition of the  $pn$ system from the normal to the superconducting phase. Also such a Hamiltonian could be used for describing the rate of double beta Fermi-type decay. Note that such a particular form of the $pn$ pairing interaction is not invariant under the rotations in the isospin space, but preserves the third component, $T_3$, of the total isospin. Indeed, the two body interaction comprises terms of isospin 0,1 and 2. 
Therefore the eigenstates of $\hat{H}$ are expected to be a mixture of components of different isospin. 
In order to have an isospin invariant Hamiltonian we have to account also for the $pp$ and $nn$ interaction. 
Due to these features we consider (\ref{ModHam})  as an illustrative example
which allows us to describe the main ingredients of the present formalism. 

\subsection{Mean field term}

The mean field energy is determined by averaging the particle number operator for each shell.
\begin{equation}
\langle N_{\alpha }\rangle =\langle BCS,N|N_{\alpha }|BCS,N\rangle . 
\end{equation}
With the interchange of the order of
the exponent operators, as described in the previous Section, this
average is transformed into
\begin{eqnarray}
\langle N_{\alpha }\rangle = Q^{-1}(N) 
\int\limits_{C}\frac{d\zeta}{2\pi i}\frac{\mathcal{P}_{\alpha}(\zeta)}{\zeta^{N+1}},
\label{NBETAaverage}
\end{eqnarray}
where 
\begin{eqnarray}
&&\hspace{-5mm}\mathcal{P}_{\alpha}(\zeta) = 2i \prod_{\gamma } (\cos ^{2}(\rho _{\gamma}/2) )^{2j_{\gamma}+ 1} \label{npnumber} \\
&&\times \frac{d}{dx} \langle 0| 
\exp ( i \sum_{\gamma } \exp (i\psi _{\gamma })\sqrt{2}\tan (\rho
_{\gamma }/2)J_{\gamma -})\nonumber\\
&&\times\exp \left( -ix \left( J_{\alpha 0}+j_{\alpha }+1/2\right) \right)   \nonumber \\
&&\times \exp ( i\zeta^{2}\sum_{\gamma } \exp (- i\psi _{\gamma })\sqrt{2}\tan (\rho
_{\gamma }/2)J_{\gamma  +} ) |0\rangle \left. {}\right| _{x=0}.\nonumber
\label{Pdef}
\end{eqnarray}
The derivative over $x$ is taken at $x=0$. Making use  of Eqs. (\ref{ORDER}), (\ref{BETA2ALPHAPM}) and (\ref{BETA2ALPHAZE}), 
$\mathcal{P}_{\alpha}(\zeta)$ can be simplified to give: 
\begin{eqnarray}
&&\mathcal{P}_{\alpha}(\zeta) = N\mathcal{G}_{\alpha}(\zeta).
\label{Pderivative}
\end{eqnarray}
The function $\mathcal{G}_{\alpha}(\zeta)$ enters the definition of $Q(N)$ and is given by Eq.~(\ref{QN}). 
Combining Eqs. (\ref{NBETAaverage}) and (\ref{Pderivative}), one obtains
\begin{eqnarray}
\langle N_{\alpha }\rangle  &=& N Q^{\alpha}(N)Q^{-1}(N). \label{averageNBETA}
\end{eqnarray}
The sum over $\alpha $ in Eq.~(\ref{averageNBETA}) gives identity $N=N$.
Using Eq.~(\ref{NUMOPE}), one can find the average of $J_{\alpha 0}$.

\subsection{Energy gap function}

Within the BCS theory the expression of the gap energy function is obtained by averaging 
the operator $\sqrt{2}\sum_{\alpha}J_{\alpha +}$ on the unprojected  BCS state. Instead, for the particle number projected  BCS formalism,  the matrix element of the mentioned operator between the states with 
$N$ and $N+2$ particles is to be calculated. We start by calculating such a matrix element for each term under summation
\begin{eqnarray}
\langle J_{\alpha +}\rangle &=&\langle BCS,N+2|J_{\alpha +}|BCS,N\rangle\nonumber\\
&=&C_{N+2}\langle BCS|J_{\alpha +}|BCS,N\rangle .
\end{eqnarray}
The BCS wave functions 
of the initial and final
states can be different ($\rho _{\alpha }\neq \rho _{\alpha}^{\prime },\,\psi _{\alpha }\neq \psi _{\alpha }^{\prime }$). 
This case is considered in Appendix A. Here, following the path described in the previous subsection,
we present results for the matrix elements diagonal in the BCS angles:
\begin{eqnarray}
\langle J_{ \alpha  +}\rangle &=& -i Q^{-1/2}(N+2)Q^{-1/2}(N)\frac{1}{2\sqrt{2}} (N+2)\nonumber \\
&\times& \exp (i\psi _{\alpha })\cot (\rho _{\alpha }/2)Q^{\alpha }(N+2). \label{BAA}
\end{eqnarray}
Alternatively one may express the above matrix elements as a polynomial in $\tan (\rho/2)$ by using a power expansion in $\zeta$ for the integrand, as explained already before. The result is:
\begin{eqnarray}
\langle J_{\alpha +}\rangle &=&iQ^{-1/2}(N+2)Q^{-1/2}(N)(2j_{\alpha }+1)
\frac{1}{2\sqrt{2}}\\ &\times&\exp (-i\psi _{\alpha })
\sum_{n=0}^{N}(-)^{n}\tan^{2n+1}(\rho _{\alpha }/2)Q(N-2n).\nonumber
\end{eqnarray} 

The matrix element of $J_{\alpha  - }$ can easily be found by complex conjugation
\begin{eqnarray}
\langle J_{\alpha  - }\rangle &=& \langle BCS,N|J_{\alpha  - }|BCS, N+2\rangle \nonumber\\
&=&- \langle J_{\alpha  +}\rangle ^*.
\end{eqnarray}
This quantity multiplied by $\sqrt{2}$ defines the spectroscopic factor for a pair $\alpha$ of states which 
could be measured in a deuteron transfer reaction.
If the two body interaction strength is state independent and equal to $-{G}/{4}$ the  pairing interaction 
term resembles the pairing interaction for alike nucleons. In virtue of the particle number conservation, 
the average of $G\sqrt{2}\sum_{\alpha}J_{\alpha +}$ is equal to zero. By analogy with the case of alike nucleon pairing, we call the sum 
\begin{equation}
\Delta^{(N)}_{pn} = G \sum_{\alpha}\langle BCS,N+2|c^{\dagger}_{\alpha m} d^{\dagger}_{\alpha \tilde{m}}|BCS, N \rangle ,
\label{gappr}
\end{equation}
the gap parameter for the $N$-nucleon system, which might be a good definition at least in the limit of large $N$.

\subsection{Interaction energy}

Now, consider the proton-neutron ($pn$) pairing interaction with the generic term:

\begin{equation}
\langle J_{ \alpha  +}J_{ - \beta}\rangle =\langle BCS,N|J_{ \alpha  +}J_{\beta -}|BCS,N\rangle . 
\end{equation}
First, we lift the operators $J_{ \alpha  +}$ and $J_{  \beta -}$ to the
arguments of exponents by introducing two derivatives over $x$ and $y$, evaluated in origin: 
\begin{eqnarray}
\langle J_{ \alpha  +}J_{\beta -}\rangle =Q^{-1}(N) \int\limits_{C} \frac{d\zeta}{2\pi i}\frac{\mathcal{P}_{\alpha \beta}(\zeta)}{\zeta^{N+1}}, \nonumber 
\end{eqnarray}
where 
\begin{eqnarray}
&&\hspace{-5mm}\mathcal{P}_{\alpha \beta}(\zeta) =- \prod_{\gamma } (\cos ^{2}(\rho _{\gamma}/2) )^{2j_{\gamma}+ 1} \\
&\times& \frac{d}{dx} \frac{d}{dy}\langle 0| \exp ( i \sum_{\gamma }\exp (i\psi _{\gamma})\sqrt{2}
\tan (\rho _{\gamma}/2)J_{\gamma-} )\nonumber\\ 
&\times&\exp (-ixJ_{ \alpha  +}) \exp ( -iyJ_{ - \beta}) \nonumber\\
&\times&\exp (
i\zeta^{2}\sum_{\delta } \exp (-i\psi _{\delta })\sqrt{2}\tan (\rho _{\delta }/2)J_{\delta +}) |0\rangle \left. {}\right| _{x=y=0}.\nonumber
\end{eqnarray}
The product of four exponents is then reordered, using the results of the Sect. II.


The final expressions for $\alpha \neq \beta $ and $\alpha = \beta $ become
\begin{widetext}
\begin{eqnarray}
\langle J_{ \alpha  +} J_{ \beta  -}\rangle &=& Q^{-1}(N)\frac{1}{4} N((2j_{\beta }+1)Q^{\alpha }(N)-(2j_{\alpha }+1)Q^{\beta}(N))
\frac{\exp (i\psi _{\beta}- i\psi _{\alpha })\tan (\rho _{\beta }/2)\tan (\rho _{\alpha }/2)}
{\tan ^2(\rho _{\beta }/2) - \tan ^2(\rho _{\alpha }/2)}, \\
\langle J_{ \alpha  +} J_{ \alpha  -}\rangle&=& - Q^{-1}(N)( \frac{N}{2} Q^{\alpha }(N) + (j_{\alpha }+1/2)
\sum_{n=1}^{N/2} (-)^{n+1} \tan ^{2n} (\rho _{\alpha }/2) (2nj_{\alpha } -1)Q(N-2n)). 
  \label{DIAG}
\end{eqnarray}
\end{widetext}

In the standard  BCS formalism, according to Eq. (3.6) the occupation probabilities for protons and for neutrons associated to the single particle  state $|\alpha\rangle$ are equal to each other.
\begin{equation}
|V^p_{\alpha}|^2=|V^n_{\alpha}|^2.
\end{equation}
Within a particle number projected formalism, the occupation probability is different from that defined above. Indeed, here the occupation probability for  the pair of $\alpha$ states is given by:
\begin{equation}
-2\langle J_{ \alpha  +}J_{ \alpha  -}\rangle \neq  \sin ^{2} (\rho _{\alpha }/2).
\end{equation}

\renewcommand{\theequation}{6.\arabic{equation}}
\setcounter{equation}{0}
\label{sec:level6}
\section{One and two shells cases}
In this Section we shall focus on the pairing Hamiltonian
\begin{eqnarray}
{\hat{H}}&=&\sum_{\alpha m}^{}\left(\epsilon_{\alpha}-\lambda \right)\left(c^{\dagger}_{\alpha m}c_{\alpha m}+
d^{\dagger}_{\alpha m}d_{\alpha m}\right)\nonumber\\
&-&G\sum_{\alpha,\beta;m,n} c^{\dagger}_{\alpha m}d^{\dagger}_{\alpha \tilde{m}}d_{\beta \tilde{n}}c_{\beta n}.
\label{pairHam}
\end{eqnarray}
In the special situation when the single particle space occupied by protons and neutrons is restricted to a single $j$, 
the energy for $N$ nucleons calculated  as the average of $\hat{H}$ corresponding to $\lambda=0$ with the $N$-projected state defined before,
is a quantity growing quadratically with $N$ and depending neither on $\rho$ nor on $\psi$. In this respect, one may say that the $N$-projected state does not exhibit a superconducting character. 
However, the spectroscopic factors defined by Eq.~(5.10) can be calculated if $\rho$ and $\psi$ 
are determined within the standard BCS. In that case, 
the spectroscopic factor is readily obtained once the functions $Q(N)$ are calculated. 

After some algebraic manipulations one finds
a Bernoulli  distribution
\begin{equation}
Q(N)= 
\left( 
\begin{array}{c}
2j+1 \\ 
N/2
\end{array}
\right) 
p^{N/2} (1 - p)^{2j + 1 - N/2}
\label{oneshellQ}
\end{equation}
with probability $p = \sin^2 (\rho/2)$, that can be written, in fact, immediately starting from the integral form (\ref{intQ}) 
or using combinatorial arguments to calculate the norm of the particle number projected BCS states.

Let us consider now that a number of nucleons, $N$, are distributed among two single particle states whose quantum 
numbers are specified through their angular momenta $j_1$ and $j_2$ respectively. 
The potential entering Eq.~(\ref{ModHam}), now has the form 
\begin{equation}
\mathcal{V}_{\alpha \beta} = 2G.
\label{pote}
\end{equation}
The corresponding energies of the shells
are denoted by $\epsilon_1$ and $ \epsilon_2$, respectively. 

\subsection{Standard BCS}

If one neglect the renormalization of the single particle energies due to the pairing interaction, 
the  energy of the constrained system of $N$ nucleons has the expression:
\begin{equation}
E'(N)=\sum_{{\alpha}}2(2j_{\alpha}+1)(\epsilon_{\alpha}-\lambda)|V_{\alpha}|^2-\frac{|\Delta|^2}{G}, 
\label{BCSEN}
\end{equation}
with
\begin{eqnarray}
\Delta &=& \frac{G}{2} \sum_{\alpha}2(2j_{\alpha}+1)U_{\alpha} V_{\alpha},\label{DBCS}\\
N&=&\sum_{{\alpha}}2(2j_{\alpha}+1)|V_{{\alpha}}|^2.                    \label{NBCS}
\end{eqnarray}

The condition of minimum energy as a function of the BCS angles can be used to express the BCS angles in terms of the energy gap $\Delta$ and the parameter $\lambda$:
\begin{eqnarray}
\left(\begin{matrix}|V_{\alpha}|^2\cr U_{\alpha}^2\end{matrix}\right)= \frac{1}{2} (1 \mp 
\frac{\epsilon_{\alpha}-\lambda}{\sqrt{(\epsilon_{\alpha}-\lambda)^2+|\Delta|^2}}).
\end{eqnarray}
The energy gap and $\lambda$ can be found from 
the self-consistency condition (\ref{DBCS}) and
the particle number constraint (\ref{NBCS}):
\begin{eqnarray}
\sum_{\alpha} \frac{(2j_{\alpha}+1)(\epsilon_{\alpha}-\lambda)}{\sqrt{(\epsilon_{\alpha}-\lambda)^2+|\Delta|^2}} &=&
\sum_{\alpha}(2j_{\alpha}+1) - N, \nonumber\\
\frac{G}{2} \sum_{\alpha} \frac{2j_{\alpha}+1}{\sqrt{(\epsilon_{\alpha}-\lambda)^2+|\Delta|^2}} &=& 1.
\label{BCSEQ}
\end{eqnarray}

The variational problem also constrains the phases. Denoting by $\varphi$ the phase of the gap parameter
\begin{equation}
\Delta = |\Delta|e^{i\varphi},
\end{equation} 
one successively find $\arg V_{\alpha} = \arg \Delta$ or, using the relation (\ref{UandV}),
\begin{eqnarray}
\psi_{\alpha} + \frac{\pi}{2} &=& \varphi.
\end{eqnarray}
The average energy (\ref{BCSEN}) is independent of the phase factor of $V_{\alpha}$. The absolute scale of the phases, therefore, is not determined.
 
If the BCS equations admit nontrivial solutions, the system is, by definition, in a superconducting phase, its energy being calculated by 
Eq.~(\ref{BCSEN}). 

In the case of a single shell, solution of the above equations has the form
\begin{eqnarray*}
\epsilon _{1}-\lambda  &=&\frac{1}{2} G (2j+1-N), \\
\Delta ^{2} &=& \frac{1}{4} G^{2}N(4j+2-N), \\
V^2_j&=&\frac{N}{2(2j+1)},\\
E'(N) &=& -\frac{G}{4}N^2.
\end{eqnarray*}
The quasiparticle energy is equal to $\frac{G}{2}(2j+1)$ while the system energy is obtained by subtracting from $E'(N)$ the contribution of the constraint term:
\begin{equation}
E_{gs}(N)=\epsilon_1N-\frac{1}{4}GN(4j+2-N).
\label{ENunproj}
\end{equation}
In the BCS theory, therefore, the superconducting state exists for any number of particles. This conclusion, however, is not supported in the particle number projected BCS theory. 
It is not difficult to see that the average energy corresponding to the $N$-projected BCS state of is independent of the BCS angles and phases, and equal to
\begin{equation}
E(N) = \epsilon_1N -  \frac{1}{4}GN (4j + 4 - N) .
\label{ENproj}
\end{equation}
This result can be obtained with the general formulas of the previous Sections, where $\lambda$ is set equal to zero but also by exploiting the fact 
 that after  projecting the nucleon total number, in the BCS wave function survives only the component $\sim (J_{+})^{N/2}|0\rangle$. The average value of the Hamiltonian $\hat{H}$, corresponding to $\lambda=0$, 
for this component provides (\ref{ENproj}). 
Note that the  the unprojected BCS state is higher in energy than and the $N$-projected BCS state.

Moreover system defined by the Hamiltonian (\ref{pairHam}) restricted to a single shell is exactly solvable, since the components of the Hamiltonian 
are expressed through the Casimir operator of the quasi-rotation group,$\hat{J}_{\alpha}^2$, and the quasi-spin projection on z-axis,$\hat{J}_{\alpha~0}$ . In our case, the expression of (\ref{ENproj}) appears to be just the eigenvalue of the pairing Hamiltonian, while the $N$-projected state is the corresponding eigenfunction.

\subsection{$N$-projected BCS}
Here we consider the case of two shells calculations. For what follows it is useful to write
the expression of the Q function in a more compact form.
The reciprocal norm squared can be written as:
\begin{eqnarray}
Q(2)&=&T_2Q(0),\\
Q(4)&=&\frac{1}{2}(T_2^2-T_4)Q(0),\nonumber\\
Q(6)&=&\frac{1}{6}\left(T_2^3-3T_2T_4+2T_6\right)Q(0),\nonumber\\
Q(8)&=&\frac{1}{24}\left(T_2^4-6T_2^2T_4+8T_2T_6+3T_4^2-6T_8\right)Q(0),\nonumber\\
Q(10)&=&\frac{1}{5}\sum_{k=1}^{5}T_{2k}Q(10-2k)
\end{eqnarray}
where
\begin{eqnarray*}
Q(0)&=& (\cos^2 (\rho_{1}/2) )^{2j_1+1}   (\cos^2 (\rho_{2}/2))^{2j_2+1}, \\
T_n &=&(2j_1+1) \tan^{n}(\rho_{1}/2) + (2j_2+1)\tan^{n}(\rho_{2}/2).
\end{eqnarray*}

The terms $Q^{\alpha}(N)$ specified by (\ref{QNrecursion1}) acquire also compact forms:
\begin{eqnarray}
Q^{{\alpha}}(2)&=&F_{\alpha 2}Q(0),\\
Q^{{\alpha}}(4)&=&\frac{1}{2}Q(0)(T_2 F_{\alpha 2}- F_{\alpha 4}),\nonumber\\
Q^{{\alpha}}(6)&=&\frac{1}{3}Q(0)(\frac{1}{2}(T_2^2-T_4)F_{\alpha 2} - T_2F_{\alpha 4}+ 
F_{\alpha 6}),\nonumber\\
Q^{{\alpha}}(8) &=& \frac{1}{4}Q(0)
(\frac{1}{6}(T_2^3-3T_2T_4+2T_6)F_{\alpha 2}\nonumber\\
&-&\frac{1}{2}(T_2^2-T_4)F_{\alpha 4} + T_2F_{\alpha 6}-F_{\alpha 8}),\nonumber\\
Q^{\alpha}(10)&=&\sum_{p=1}^{5}F_{k,2p}(-1)^{p+1}Q(10-2p).
\end{eqnarray}
where
\[
F_{\alpha n} = (2j_{\alpha}+1)\tan^n (\rho_{\alpha}/2).
\]
Using these partial results in connection with the matrix elements (5.6), (5.13) and (5.14) the system energy calculated as the average value of the pairing Hamiltonian $\hat{H}$ corresponding to the $N$-projected BCS state, is readily obtained.

\begin{figure}
\begin{center}
\includegraphics[height=8cm,width=0.5\textwidth]{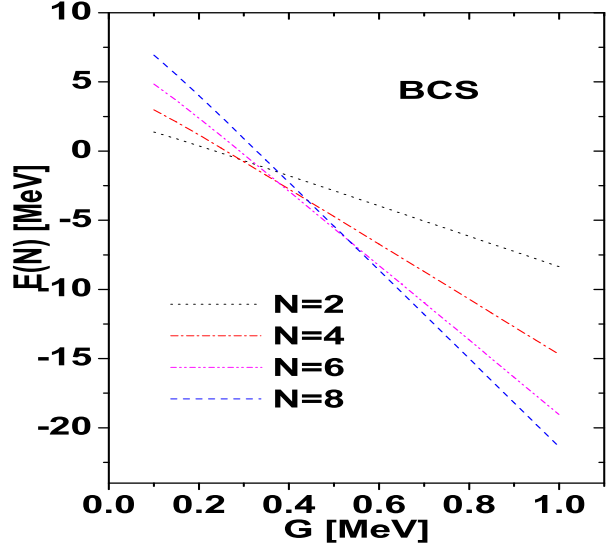}
\end{center}
\caption{The energy of a system with N nucleons ($N=2,4,6,8$ ), obtained within the BCS formalism, is plotted as  function of the pairing interaction strength, G. }    
\end{figure}
\begin{figure}
\begin{center}
\includegraphics[height=8cm,width=0.5\textwidth]{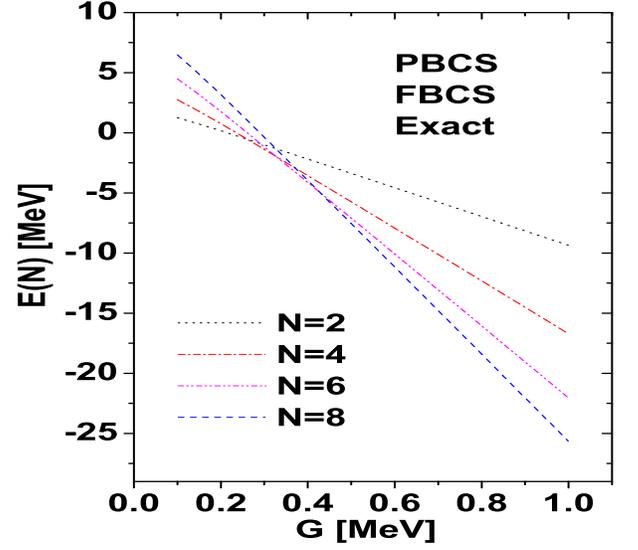}
\end{center}
\caption{The energy of a system with N nucleons ($N=2,4,6,8$ ),obtained within the PBCS formalism, is plotted as  function of the pairing interaction strength, G. Energies predicted by the PBCS, the FBCS and diagonalization (Exact) are identical. This is indicated by assigning to the given curves all three labels: PBCS,FBCS, Exact. }    
\end{figure}

\begin{figure}
\begin{center}
\includegraphics[height=8cm,width=0.5\textwidth]{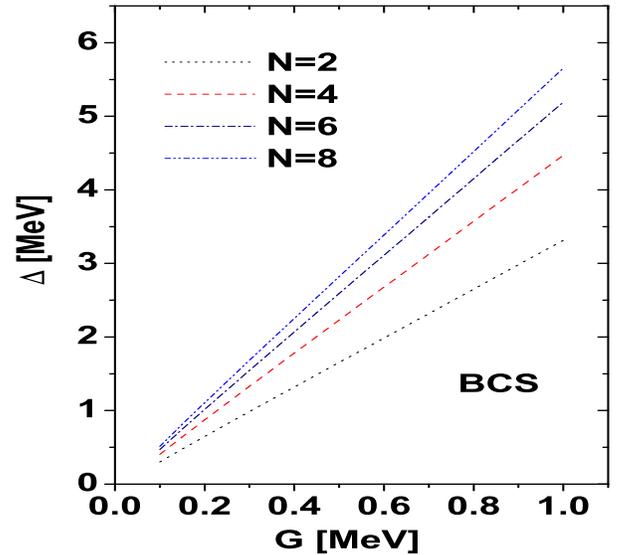}
\end{center}
\caption{The gap energy for a system with N nucleons ($N=2,4,6,8$) is plotted as  function of the pairing interaction strength, G, within the standard BCS formalism. }    
\end{figure}
\begin{figure}
\begin{center}
\includegraphics[height=8cm,width=0.5\textwidth]{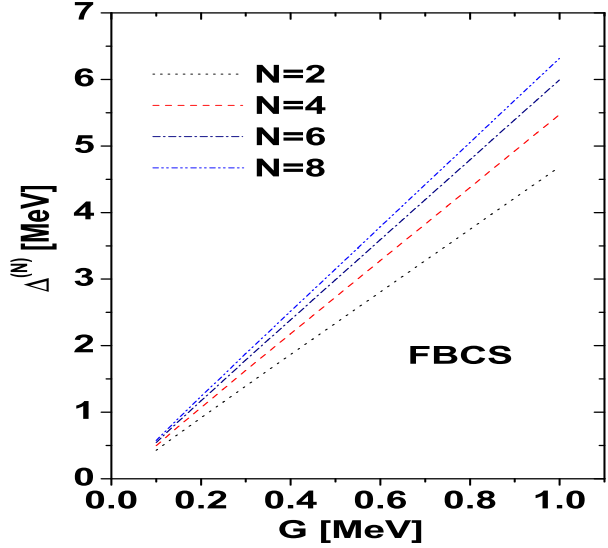}
\end{center}
\caption{The gap energy for a system with N nucleons ($N=2,4,6,8$) is plotted as  function of the pairing interaction strength, G, within the standard FBCS formalism. The gap function is defined by Eq. (\ref{gappr}).}    
\end{figure}
\begin{table}
\begin{tabular}{|c|cc|cc|}
\hline
     &\multicolumn{2}{c|}{G=0.1 MeV}&\multicolumn{2}{c|}{G=1.0 MeV}\\
\cline{2-5}
      & N=2 & N=8& N=2& N=8\\
\hline
$E_{BCS}$&1.37573 & 6.93772 & -8.35069 & -21.38316 \\
$\Delta$ &0.30183 & 0.51425 &  3.31395 &   5.65241  \\
$E_{PBCS}$&1.25987 &6.48216 &-9.35235 & -25.38779      \\
$E_{FBCS}$&1.25969 & 6.48213 & -9.35235 &-25.38779    \\
$\Delta^{(N)}$&0.4292 &0.5753 & 4.6 866 & 6.3196 \\
\hline
\end{tabular}
\caption{The ground state energy provided by the BCS, PBCS and FBCS formalism, respectively
, given in MeV, are listed for two values of the nucleon total number and two values of the pairing strength, G.}
\end{table}

Making use of the results obtained so far one can calculate the ground state energies as well as the energy gap. Calculations were successively performed for the standard BCS, the projection after variation approach, PBCS, and the projection before variation formalism , FBCS. Also, the exact eigenvalues of $\hat{H}$ have been obtained by diagonalization.
The input data in our calculation are:

\begin{eqnarray}
j_1&=&\frac{3}{2},\;\;j_2=\frac{7}{2}, \nonumber\\
\epsilon_1&=&1~\mathrm{MeV},\;\;\epsilon_2=1.5~\mathrm{MeV}.
\end{eqnarray}
For a given $N$ ($=2,4,6,8$) we solved the BCS equations (\ref{BCSEQ}) and then calculated the ground state energy
(\ref{BCSEN}). Results for energies are plotted, in Fig.~1, as function of the pairing interaction strength. 
The uncorrelated system has the energy
\begin{equation} 
E_{normal}=N\epsilon_1.
\end{equation} 

Alternatively, with the 
parameters $\rho_{1}, \rho_{2}, \psi_{1}, \psi_{2}$ determined by the standard BCS approach, 
the system energy was calculated with the general expression provided by the present formalism with the $N$-projected state. We conventionally call this approach as projected BCS (PBCS). The resulting energies are represented in Fig.2 as functions of $G$. 

For each $N$, further,  the energy $E(N)$ given by the $N$-projection formalism was minimized with respect to the parameters involved in the 
trial function, i.e. $\rho_{1}, \rho_{2}, \psi_{1}, \psi_{2}$.
It turns out that the ground state energies provided by PBCS, FBCS and diagonalization are equal to each others for any $G$ in the range of 0.1-1.0 MeV. The common values were used in Fig.2 showing the energy dependence on the pairing strength parameter.

The two figures mentioned before exhibit some common features. Energies are decreasing functions of G. They show a linear dependence on G with the slope depending on the particle total number, $N$. For low values of G, energies are increasing functions of N, while for large pairing strength they are decreasing with $N$. In the four approaches, the transition from one energy ordering to another is taking place for different values of G. The largest critical $G$ is met for the standard BCS. To get a better view on the quantitative energy split due to its $N$-dependence we collected the energy values for the extreme values of $N$, in Table I.
From there one can see that the energy ordering for a given $G$ is not depending on $N$. Indeed, for
$G=0.1 \;MeV$ the ordering for $N=2$ and $N=8$ are:
\begin{equation}
E_{normal}>E_{BCS}>E_{PBCS}>E_{FBCS},\;\;N=2,8,\nonumber\\
\end{equation}
while for $G=1.0\; MeV$ the energy order is as:
\begin{equation}
E_{normal}>E_{BCS}>E_{PBCS}=E_{FBCS},\;\;N=2,8.\nonumber\\
\end{equation}
It is remarkable the fact that energies provided by the PBCS and FBCS approaches are practically the same. Only very tiny differences are noticed for G=0.1 MeV, which might be caused by the numerical procedures adopted for the two cases.

Due to the above mentioned ordering for the ground state energies provided by different formalism we may conclude that for all cases the superconducting phase is achieved for any $G\geq 0.1$.
Another observable considered in our study is the energy gap obtained within the standard BCS and
the FBCS formalism, respectively. The results obtained for a fixed $N$ were plotted as function of $G$ in figure 3 and 4, respectively. These figures show a linear dependence of both gaps on G.
The split of gaps due to their $N$ dependence is larger for $\Delta$ than for $\Delta^{(N)}$.
Indeed, according to the data listed in Table I, $\Delta(N=8)-\Delta(N=2)$ is equal to 0.212 MeV and 2.338 MeV for $G$ equal to 0.1 MeV and 1.0 MeV respectively, while $\Delta^{(8)}-\Delta{^(2)}$ for $G=0.1; 1.$MeV, amounts to 0.146 MeV and 1.633 MeV, respectively.
Note that for a given set of $N$ and $G$ we have $\Delta >\Delta^{(N)}$. Since the energy gap might be looked at as a measure of superconductivity one may expect that the superconductivity effects are more pronounced in the PBCS and FBCS than in the BCS.

The Hamiltonian $\hat{H}$ can be diagonalized in a basis of definite number of particles, $N$
\begin{eqnarray}
|N_1,N_2\rangle&=&C_{N_1N_2}(c^{\dagger}_1d^{\dagger}_{\tilde{1}})^{N_1/2}(c^{\dagger}_2
d^{\dagger}_{\tilde{2}})^{N_2/2}|0\rangle,\nonumber\\
N_1+N_2&=&N.
\end{eqnarray}
with $C_{N_1N_2}$ standing for the normalization constant.
The matrix elements of $H$ in this basis have simple expressions:
\begin{widetext}
\begin{eqnarray}
&&\langle N_1,N_2|H|N_1,N_2\rangle=\epsilon_1N_1+\epsilon_2N_2  
-\frac{G}{4}\left(N_1(4j_1+4-N_1)+N_2(4j_2+4-N_2)\right),\nonumber\\
&&\langle N_1+2,N_2-2|H|N_1,N_2\rangle =-\frac{G}{4}\left[(N_1+2)N_2(4j_1+2-N_1)(4j_2+4-N_2)\right]^{1/2},\nonumber\\
&&\langle N_1-2, N_2+2|H|N_1,N_2\rangle =-\frac{G}{4}\left[N_1(N_2+2)(4j_1+4-N_1)(4j_2+2-N_2)\right]^{1/2}.
\end{eqnarray}
\end{widetext}

For each $N$ and a fixed $G$, we diagonalized the above matrix and depicted the lowest eigenvalues.
Further, these were compared with the energies provided by the FBCS  and the PBCS formalisms. In this way we found out that the three sets of energies are identical. This is a nice example when the solution of two variational principle equations reproduce the exact ground state energy. However,  in a realistic single particle space and, moreover, when an isospin invariant Hamiltonian is instead considered, this feature does not necessarily show up. 

\renewcommand{\theequation}{6.\arabic{equation}}
\setcounter{equation}{0}
\section{Conclusions}

Many interesting properties of the $pn$ pairing have been derived using for the BCS function the form (3.1). After making use of the factorization described in Section I, one finds out that the function (3.1) for a fixed $\alpha$ represents, the coherent state of the SU(2) group generated by $J_{\alpha\mu}$. Our attention was focused on the nucleon number projected BCS. Since any matrix element can be expressed in terms of the norms of the involved states, we started with the norm calculation. One of the main results of the present paper is the recursive formula for these norms. By this equation the norm of a projected state with N particles is related with the norms of the $N'$-projected states with $N'<N$. For pairing interaction of alike nucleons, a similar recursion formula was obtained in Ref.~(\cite{DMP}). The difference between the two recursion formulas consists of that there the recursion is operating in two dimensions, there are two indices which are iterated, while here only one index is involved in the recursion. 

To prove the usefulness of the obtained recursion formula several matrix elements have been evaluated. The one for the proton-neutron pairing operator of a given shell, is interpreted as
spectroscopic factor for a deuteron transfer reaction. Being guided by the analogy with the
alike nucleon pairing we defined a quantity which might be a measure for the energy gap in the particle number projected picture. Also, we calculated the occupation probability for a given state with a proton-neutron pair. In Appendix A the matrix element involved in the width of an $\alpha$ decay process is analytically expressed. Also, the matrix element for the two body proton-neutron interaction in the particle-particle channel between the states associated with the mother and daughter nuclei involved in a double beta Fermi decay, are obtained in a compact form.  

Using a Hamiltonian including a mean field term and a proton-neutron pairing interaction we calculated the system energy as function of the particle number N and
the parameters $\rho_{\alpha},\psi_{\alpha}$ defining the unprojected BCS wave function. 
Since many features of the paired system can be found also in a restricted single particle space we discussed the simple cases of one and two single particle states. Since the energy associated to a single $j$ and the $N$-projected picture, is constant with respect to the BCS parameters we concluded that for this case a superconducting phase cannot be reached. However, the case of two j  is suitable for studying the pairing properties both in the standard BCS and projected BCS formalism. In the later situation we considered both cases when the variation is performed before and after projection.
In the two level situation we have proved that the ground state energy provided by the PBCS, FBCS are the same and moreover equal to the exact ground state energy obtained by diagonalization.
It is an open question whether this feature is caused by the  restricted single particle space or by the particular choice o the model many-body Hamiltonian.

One of the evident limitations of our formalism consists in the fact that the trial BCS unprojected function allows us to describe only nuclei with N=Z. This feature can be however improved by adding to (3.1) two factors accounting for the proton-proton and neutron-neutron pairing, respectively.

As an imminent project for the near future we also mention the extension of our formalism to isospin preserving Hamiltonians. In the second step of the formalism development we shall attempt to include in our study the proton-neutron $T=0$ pairing as well as the isospin projection. This plan is, in fact, a reflection of our belief that a new and powerful technical result might be decisive in unveiling new properties of the paired proton-neutron system.

{\bf Acknowledgment.} This work was partially supported by the Romanian Ministry for Education Research Youth and Sport through the CNCSIS project ID-2/5.10.2011, by the grant No. 4568.2008.2 of Leading Scientific Schools of Russian Federation, DFG project RUS 113/721/0-3, and RFBR project No. 09-02-91341.

\appendix

\renewcommand{\theequation}{A.\arabic{equation}}
\setcounter{equation}{0}
\label{sec:levelA}
\section{Matrix elements of different final and initial particle number projected BCS states}

In various applications matrix elements of operators between two different
BCS states are frequently encountered. In the case of projected BCS, such
matrix elements are calculated in the manner described in Section V. Here we
derive formulas for $\langle BCS_{f},N_{f}|O|BCS_{i},N_{i}\rangle $ for 
$O=1$, $J_{ \pm \alpha},$ $J_{\pm \alpha }J_{\pm \beta}$ and $J_{ \mp \alpha}J_{ \pm \beta}$ with different BCS
angles of the final and initial states.

We start from the diagonal in particle number overlap
\begin{eqnarray*}
\langle BCS_{f},N|BCS_{i},N\rangle  &=&Q_{f}^{-1/2}(N)\langle BCS_{f}|BCS_{i},N\rangle  \\
&=&Q_{f}^{-1/2}(N)Q_{i}^{-1/2}(N)Q_{fi}(N),
\end{eqnarray*}
where $Q_{f}(N)$ and $Q_{i}(N)$ are the diagonal $Q$-functions defined for
the final- and initial-state BCS angles $(\rho _{\alpha }^{\prime },\psi
_{\alpha }^{\prime })$ and $(\rho _{\alpha },\psi _{\alpha })$,
respectively. 
Calculations similar to those carried out in Section V give give the off-diagonal $Q$-function 
\begin{widetext}
\begin{equation}
Q_{fi}(N)= \int \limits_{C}\frac{d\zeta }{2\pi i}\frac{1}{\zeta ^{N+1}} \prod_{\alpha }\left( \cos (\rho _{\alpha }^{\prime }/2)\cos (\rho _{\alpha
}/2)+\zeta ^{2}\exp (i\psi _{\alpha }^{\prime }-i\psi _{\alpha })\sin (\rho
_{\alpha }^{\prime }/2)\sin (\rho _{\alpha }/2)\right) ^{2j_{\alpha }+1}.
\label{bound}
\end{equation}
Integrating by parts and expanding into a series in $\zeta $ or $%
1/\zeta $, we obtain the recursion for the calculation of this function: 
\begin{eqnarray}
Q_{fi}(N) &=&\sum_{\alpha }Q_{fi}^{\alpha }(N), \nonumber \\
Q_{fi}^{\alpha }(N) &=&\frac{\Omega_{\alpha }}{N}
\sum_{n=1}^{N/2}(-)^{n+1}
\left( \exp (i\psi _{\alpha }^{\prime }-i\psi _{\alpha }) \tan (\rho _{\alpha }^{\prime }/2)\tan(\rho _{\alpha }/2) \right)^{n} Q_{fi}(N-2n) \label{A2} \\
&=&\frac{\Omega_{\alpha }}{\Omega-N}\sum_{n=1}^{(\Omega -N)/2}(-)^{n+1}
\left( \exp (-i\psi _{\alpha }^{\prime }+i\psi _{\alpha })\cot(\rho _{\alpha }^{\prime }/2)\cot (\rho _{\alpha }/2) \right)^{n} Q_{fi}(N+2n). \label{A3}
\end{eqnarray}
\end{widetext}
The boundary conditions follow from (\ref{bound}) 
\begin{eqnarray*}
Q_{fi}(0) &=&\prod_{\alpha }\left( \cos (\rho _{\alpha }^{\prime }/2)\cos
(\rho _{\alpha }/2)\right) ^{2j_{\alpha }+1}, \\
Q_{fi}(\Omega ) &=&\prod_{\alpha }\left( \exp (i\psi _{\alpha }^{\prime }-i\psi _{\alpha}) \sin (\rho _{\alpha }^{\prime}/2)\sin (\rho _{\alpha }/2)\right) ^{2j_{\alpha }+1}.
\end{eqnarray*}
As in the case of the function $Q(N)$, $Q_{fi}(N)$ vanishes outside the
interval $(0,\Omega )$ and for odd $N$. In the limit where the angles in the
initial and final BCS states coincide, we recover the result for $Q(N)$. 
In particular, $Q_{ff}(N)=Q_{f}(N)$ and $Q_{ii}(N)=Q_{i}(N)$.
$Q_{fi}^{\alpha}(N)$ vanishes outside the
interval $(2,\Omega - 2)$.

Now, consider the $2$-fermion gap function 
\begin{eqnarray*}
\langle J_{ \alpha  +}\rangle _{fi} &=&\langle BCS_{f},N+2|J_{ \alpha  +}|BCS_{i},N\rangle \\
&=&Q_{f}^{-1/2}(N+2)\langle BCS_{f}|J_{ \alpha  +}|BCS_{i},N\rangle .
\end{eqnarray*}
The calculation is performed using the representation of $J_{ \alpha  +}$ in
the exponential form 
\[
J_{ \alpha  +}=-i\frac{d}{dx}\left. \exp (iJ_{ \alpha  +}x)\right| _{x=0} 
\]
and Eq. (\ref{mostimportantaverage}). A similar representation is used for the other quasi-spin components. After some algebraic manipulations, we obtain 
\begin{widetext}
\begin{eqnarray}
\langle J_{ \alpha  +}\rangle _{fi} &=& -iQ_{f}^{-1/2}(N+2)Q_{i}^{-1/2}(N)\frac{1}{2\sqrt{2}} (N+2)\exp (i\psi _{\alpha })\cot (\rho _{\alpha }/2)Q_{fi}^{\alpha }(N+2), \label{AA} \\
\langle J_{\alpha 0}\rangle _{fi} &=&Q_{f}^{-1/2}(N)Q_{i}^{-1/2}(N)(\frac{N}{2}Q_{fi}^{\alpha }(N)-(j_{\alpha }+1/2)Q_{fi}(N)). \label{AB}
\end{eqnarray}
\end{widetext}
Summing the last equation over $\alpha$ one arrives at:
\[
2\sum_{\alpha }\langle J_{0 \alpha }\rangle _{fi} = Q_{f}^{-1/2}(N)Q_{i}^{-1/2}(N)Q_{fi}(N)(N-\frac{1}{2}\Omega),
\]
which is consistent with the definition for  particles number operator, Eq.~(\ref{NUMOPE}).
The expression for $\langle J_{\alpha  -}\rangle _{fi}$ is obtained by complex conjugation (\ref{AA}) .

Consider now the $4$-fermion gap function 
\[
\langle J_{ \alpha  +}J_{ \beta  +}\rangle _{fi}=\langle BCS_{f},N+4|J_{\alpha
+}J_{ \beta  +}|BCS_{i},N\rangle .
\]
This matrix element represents the width of the $\alpha$ decay  of the mother nucleus in the state $|BCS_f,N+1\rangle$ to the daughter nucleus in the state $|BCS_i,N_f\rangle$. Also, based on this 
matrix elements one may define the spectroscopic factor of a reaction which removes an $\alpha$ particle from the states $\alpha$ and $\beta$. In other words squaring this matrix element one obtains the occupation probabilities of the states $\alpha$ and $\beta$ with proton-neutron pairs. 
The calculation uses the representation 
\[
J_{ \alpha  +}J_{ \beta  +}=-\frac{d}{dx}\frac{d}{dy} \left. \exp (iJ_{ \alpha  +}x+iJ_{ \beta  +}y)\right| _{x=y=0}
\]
and the method of Section II. For $\alpha \neq \beta ,$ we obtain 
\begin{widetext}
\begin{eqnarray}
\langle J_{ \alpha  +}J_{ \beta  +}\rangle _{fi}
&=&-Q_{f}^{-1/2}(N+4)Q_{i}^{-1/2}(N)\frac{1}{4}  (N+2)((2j_{\beta }+1)Q_{fi}^{\alpha }(N+2)-(2j_{\alpha}+1)Q_{fi}^{\beta }(N+2)) \nonumber \\
&\times& 
\frac{\exp (i\psi _{\beta}^{\prime }+i\psi _{\alpha }^{\prime }) \tan (\rho _{\alpha }^{\prime }/2)\tan(\rho _{\beta }^{\prime }/2) }
{\exp (i\psi _{\alpha  }^{\prime }-i\psi _{\alpha  }) \tan (\rho_{\alpha  }^{\prime }/2)\tan (\rho_{\alpha  }/2)
-\exp (i\psi _{\beta }^{\prime }-i\psi _{\beta }) \tan (\rho_{\beta }^{\prime }/2)\tan (\rho_{\beta }/2)} .
\label{AC}
\end{eqnarray}
Similar expression is obtained for the matrix element
\begin{eqnarray}
\langle J_{ \alpha  +} J_{ \beta  -}\rangle _{fi} &=&\langle BCS_{f},N|J_{ \alpha  +} J_{ \beta  -}|BCS_{i},N\rangle \nonumber \\
&=&-Q_{f}^{-1/2}(N)Q_{i}^{-1/2}(N)\frac{1}{4} N((2j_{\beta }+1)Q_{fi}^{\alpha }(N)-(2j_{\alpha }+1)Q_{fi}^{\beta}(N)) \nonumber \\
&\times& \frac{\exp (i\psi _{\beta}^{\prime }-i\psi _{\alpha })\tan (\rho _{\beta }^{\prime }/2)\tan (\rho _{\alpha }/2)}
{\exp(i\psi _{\alpha }^{\prime }-i\psi _{\alpha }) \tan (\rho _{\alpha }^{\prime }/2)\tan (\rho _{\alpha }/2)
-\exp (i\psi _{\beta }^{\prime }-i\psi _{\beta})\tan (\rho _{\beta }^{\prime }/2)\tan (\rho _{\beta }/2)}.
\end{eqnarray}
We may ask ourself how this situation may appear?  The answer is offered by the double beta decay
with/without neutrinos in the final state. Indeed, in such a process the nucleus $|N,Z\rangle$ goes
to the nucleus $|N-2,Z+2\rangle$, two electrons and two/zero anti-neutrinos. The mentioned states are described by different sets of BCS angles and phases but have equal number of nucleons.
As a matter of fact the two body interaction whose matrix element is calculated is nothing else but the particle-particle interaction of the Fermi type. These comments prove that to calculate such a matrix element is an important step in describing some important physical processes.

The average (\ref{AC}) is symmetric under interchange of the indices and 
$\langle J_{ \alpha  +} J_{ \beta  -}\rangle _{fi} =\langle J_{ \beta  -} J_{ \alpha  +} \rangle _{fi}  $. 
Complex conjugation of (\ref{AC}) allows to find $\langle J_{\alpha  - } J_{\beta  -}\rangle _{fi} $.

When shells are the same, the result is as follows
\begin{eqnarray}
\langle J_{ \alpha  +}J_{ \alpha  +}\rangle_{fi}  &=&-Q_{f}^{-1/2}(N+4)Q_{f}^{-1/2}(N)\frac{j_{\alpha }}{2}\exp (i\psi _{\alpha }+i\psi _{\alpha }^{\prime })\tan
(\rho _{\alpha }^{\prime }/2)\cot (\rho _{\alpha }/2) \nonumber  \\
&\times& \sum_{n=0}^{N/2}(-)^{n}(\exp (i\psi _{\alpha }^{\prime }-i\psi_{\alpha })\tan (\rho _{\alpha }^{\prime }/2)\tan (\rho _{\alpha}/2))^{n}
(N+2-2n)Q_{fi}^{\alpha }(N+2-2n) \label{A4} \\
&=& -Q_{f}^{-1/2}(N+4)Q_{i}^{-1/2}(N) j_{\alpha }(2j_{\alpha }+1) \exp (2i\psi _{\alpha }^{\prime })\tan ^{2}(\rho _{\alpha }^{\prime }/2) \nonumber \\
&\times& \sum_{n=0}^{N/2}(-)^{n}(\exp (i\psi _{\alpha }^{\prime }-i\psi _{\alpha })\tan (\rho _{\alpha }^{\prime }/2)\tan (\rho _{\alpha }/2))^{n}(n+1)Q_{fi}(N-2n),
\label{A5}\\
\langle J_{ \alpha  +}J_{ \alpha  -}\rangle_{fi}&=& - Q_{f}^{-1/2}(N)Q_{i}^{-1/2}(N)( \frac{N}{2}Q_{fi}^{\alpha }(N)\nonumber \\ 
&+&  (j_{\alpha }+1/2)%
\sum_{n=1}^{N/2} (-)^{n+1}(\exp (i\psi _{\alpha }^{\prime }-i\psi _{\alpha
})\tan (\rho _{\alpha }^{\prime }/2)\tan (\rho _{\alpha }/2))^{n}(2nj_{\alpha } - 1)Q_{fi}(N-2n)). 
\label{A6}
\end{eqnarray}
\end{widetext}
The representations (\ref{A4}) and (\ref{A5}) are equivalent.

The formalism allows also to provide an expression for the average as a sum
over the quantum numbers $\alpha $ for a fixed number of the particles as well. However, the number of shells is
usually much larger than $N$, and the ratio between $\Omega $ and $N$
indicates the accuracy of the solution of the variational problem. The
higher the ratio, the higher the accuracy. For this reason, the summation
over the number of particles is easier from the computational point of view,
and therefore preferable. Also, the recursive computation of the function $Q_{fi}(N)$ starting from small numbers of particles is more simple.

The function $Q_{fi}(N)$ can be written in the form \cite{RaMo}
\begin{widetext}
\begin{eqnarray}
Q_{fi}(N) = \sum\limits_{N_{1}+N_{2}+\ldots = N} \prod\limits_{\alpha }\frac{(\Omega_{\alpha }/2)!}{(\Omega_{\alpha }/2-N_{\alpha }/2)!(N_{\alpha }/2)!}
(\exp (-i\psi _{\alpha }^{\prime }+i\psi _{\alpha })&&\hspace{-4mm}\sin (\rho _{\alpha }^{\prime }/2)\sin (\rho _{\alpha }/2))^{N_{\alpha }/2} \nonumber \\
 \times 
(&&\hspace{-4mm}\cos (\rho _{\alpha }^{\prime }/2)\cos (\rho _{\alpha }/2))^{\Omega_{\alpha }/2-N_{\alpha }/2}.
\label{A10}
\end{eqnarray}
\end{widetext}
Here the summation is over all sets of occupation numbers $N_{\alpha}$, which gives the total number of particle $N$, the occupation numbers are even. 
In the special case of a single shell, we reproduce the equation (\ref{oneshellQ}).

\end{document}